\documentclass[runningheads]{llncs}

\usepackage[T1]{fontenc}

\usepackage{stmaryrd}      
\usepackage[colorlinks,linkcolor=blue]{hyperref}
\usepackage{graphicx,verbatim}
\usepackage{amssymb}
\usepackage{amsmath}
\usepackage{amssymb}
\usepackage{multirow}
\usepackage{tabularx}

\usepackage{url}
\usepackage[table,xcdraw]{xcolor}
\usepackage{amsmath}
\usepackage{xcolor}
\usepackage{graphicx} 
\usepackage{booktabs}
\usepackage{mathrsfs}
\usepackage[misc]{ifsym}
\usepackage{float}

\begin{document}
\title{UltraTwin: Towards Cardiac Anatomical Twin Generation from Multi-view 2D Ultrasound}
\titlerunning{UltraTwin}
\authorrunning{J. Yu et al.}

\author{
Junxuan Yu\inst{1}\thanks{These authors contributed equally.} \and  
Yaofei Duan\inst{2\star} \and 
Yuhao Huang\inst{1\star} \and 
Yu Wang\inst{5\star} \and 
Rongbo Ling\inst{1} \and 
Weihao Luo\inst{1} \and 
Ang Zhang\inst{1} \and 
Jingxian Xu\inst{1} \and 
Qiongying Ni\inst{1} \and 
Yongsong Zhou\inst{1} \and 
Binghan Li\inst{1} \and 
Haoran Dou\inst{3} \and 
Liping Liu\inst{5} \and 
Yanfen Chu\inst{6} \and 
Feng Geng\inst{7} \and 
Zhe Sheng\inst{8} \and 
Zhifeng Ding\inst{8} \and 
Dingxin Zhang\inst{8} \and 
Rui Huang\inst{9} \and 
Yuhang Zhang\inst{10} \and 
Xiaowei Xu\inst{11} \and 
Tao Tan\inst{2} \and 
Dong Ni\inst{1} \and 
Zhongshan Gou\inst{4}\textsuperscript{\Letter} \and 
Xin Yang\inst{1}\textsuperscript{\Letter}   
} 

\institute{
National-Regional Key Technology Engineering Laboratory for Medical Ultrasound, School of Biomedical Engineering, Medical School, Shenzhen University, China \and
Faculty of Applied Sciences, Macao Polytechnic University, Macao, China \and
Centre for CIMIM, Manchester University, Manchester, UK \and
Center for Cardiovascular Disease, The Affiliated Suzhou Hospital of Nanjing Medical University, Suzhou Municipal Hospital, Gusu School, Suzhou, China \and
The First Affiliated Hospital of Kunming Medical University, Kunming, China \and
The First Affiliated Hospital of Anhui Medical University North, Hefei, China \and
The Second People’s Hospital, Wuhu, China \and
The First Affiliated Hospital of Anhui Medical University, Hefei, China \and
The People's Hospital of Tongren City, Tongren, China \and
The First People's Hospital of Taicang City, Taicang, China \and
Guangdong Provincial People’s Hospital, Guangzhou, China\\
\email{gzhongshan1986@163.com}, \email{xinyang@szu.edu.cn}.
}\maketitle              

\begin{abstract}
Echocardiography is routine for cardiac examination.
However, 2D ultrasound (US) struggles with accurate metric calculation and direct observation of 3D cardiac structures.
Moreover, 3D US is limited by low resolution, small field of view and scarce availability in practice.
Constructing the cardiac anatomical twin from 2D images is promising to provide precise treatment planning and clinical quantification. 
However, it remains challenging due to the rare paired data, complex structures, and US noises.
In this study, we introduce a novel generative framework \textbf{UltraTwin}, to obtain cardiac anatomical twin from sparse multi-view 2D US. 
Our contribution is three-fold.
First, pioneered the construction of a real-world and high-quality dataset containing strictly paired multi-view 2D US and CT, and pseudo-paired data.
Second, we propose a coarse-to-fine scheme to achieve hierarchical reconstruction optimization.
Last, we introduce an implicit autoencoder for topology-aware constraints.
Extensive experiments show that UltraTwin reconstructs high-quality anatomical twins versus strong competitors. We believe it advances anatomical twin modeling for potential applications in personalized cardiac care.


\end{abstract}

\section{Introduction}
\begin{figure}[!h]
\centering
\includegraphics[width=1\textwidth]{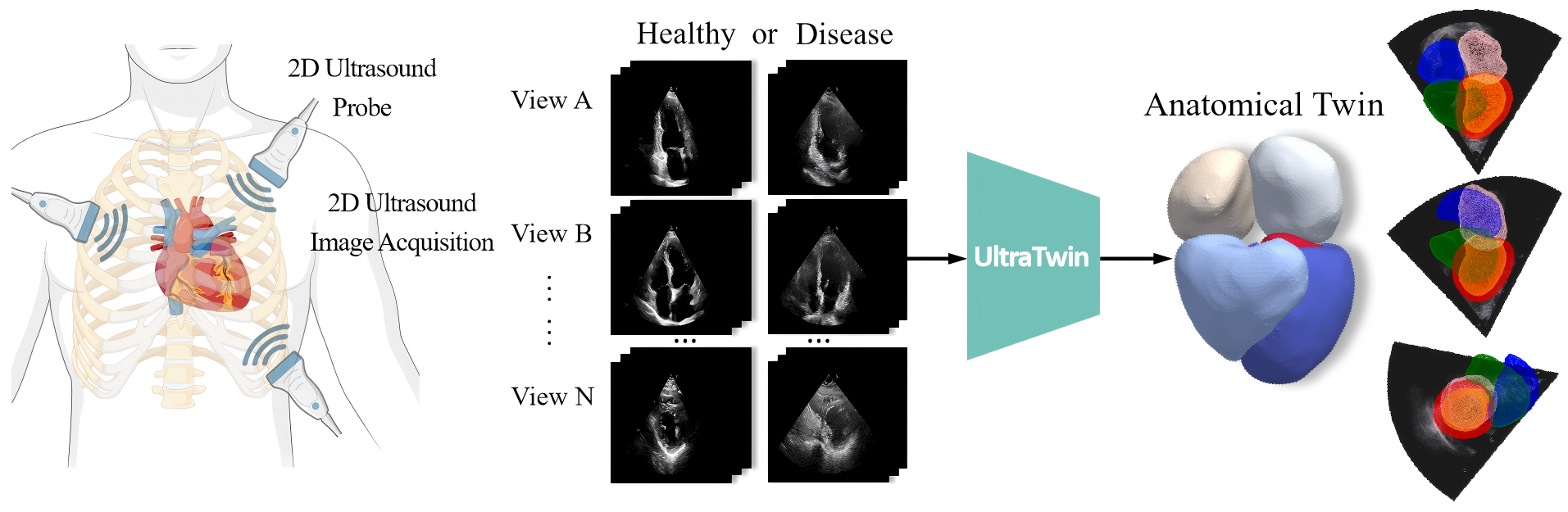}
\caption{UltraTwin Workflow: from multi-view 2D US to 3D cardiac anatomical twin.} 
\label{fig0}
\end{figure}

Accurate assessment of cardiac structure and function is vital for diagnosis, treatment planning, and surgical guidance. 
However, 2D ultrasound (US) lacks precise parameter evaluation and 3D anatomical context, bringing significant uncertainties in clinical decision-making, particularly for complex cardiac disease cases.
Additionally, 3D US is hindered by motion-induced artifacts, restricted field of view, and limited clinical availability~\cite{nelson2000sources}.
Hence, reconstructing  cardiac anatomical
twins directly from sparse 2D US is highly desired (see Fig.~\ref{fig0}).
These cardiac anatomical twins have the great potential to facilitate the personalized evaluation of cardiac function and improve treatment planning~\cite{li2024towards,thangaraj2024cardiovascular,bernard2018deep}.

Recently, several deep learning-based algorithms for cardiac reconstruction have been investigated. 
Laumer et al.~\cite{laumer2023weakly} utilized single-view US for synthetic cardiac reconstruction.
However, their approach could not differentiate between variations inherent to US imaging artifacts and actual cardiac pathologies due to the limited one-view information and lack of 3D ground truth (GT) validation. 
Stojanovsk et al.~\cite{stojanovski2022efficient} then extended the approach to multi-view reconstruction by incorporating cardiac wall masks as guidance. 
While their method alleviated the single-view problem, it seriously depends on accurate mask segmentation, leading to potential error propagation. 
To capture the complexity and diversity of real clinical cases, research~\cite{laumer20252d} attempted left ventricular reconstruction using multi-view US images. 
However, it overlooked crucial inter-chamber relationships and lacked validation with 3D GT. 
Reconstructing 3D cardiac structures from sparse 2D US views remains challenging, as multiple 3D cardiac shapes can correspond to the same set of 2D views, complicating personalized reconstruction. 
While Diffusion Models~\cite{dm} offer promising solutions through distribution modeling, their data-intensive nature poses additional challenges especially when paired 2D-3D data is scarce. 
To address the above challenges, we propose UltraTwin, the first multi-view 3D reconstruction framework based on Diffusion Transformer (DiT)~\cite{peebles2023scalable} in real-world cardiac imaging scenarios. 
We prospectively collected a real-world dataset of paired multi-view US images and corresponding CT scans for 3D modeling. 
UltraTwin encompasses three key designs: 
(1) Cardiac modeling pretraining with a novel pseudo-paired data generation strategy for robust data augmentation, enabling effective learning from limited clinical data. 
(2) A coarse-to-fine reconstruction pipeline that progressively refines cardiac structures through template-to-anatomy hierarchical feature learning, ensuring both global anatomical consistency and local anisotropic information extraction. 
(3) Implicit topology perception constraints that leverage cardiac structure priors to reduce noise in 2D US projections and enhance the anatomical plausibility of 3D reconstructions. 
Comprehensive validation on the large dataset demonstrates UltraTwin's superior performance across multiple quantitative metrics, showing promising potential for clinical cardiac chamber volume measurements.

\section{Methodology}
\begin{figure}[H]
\centering
\includegraphics[width=1\textwidth]{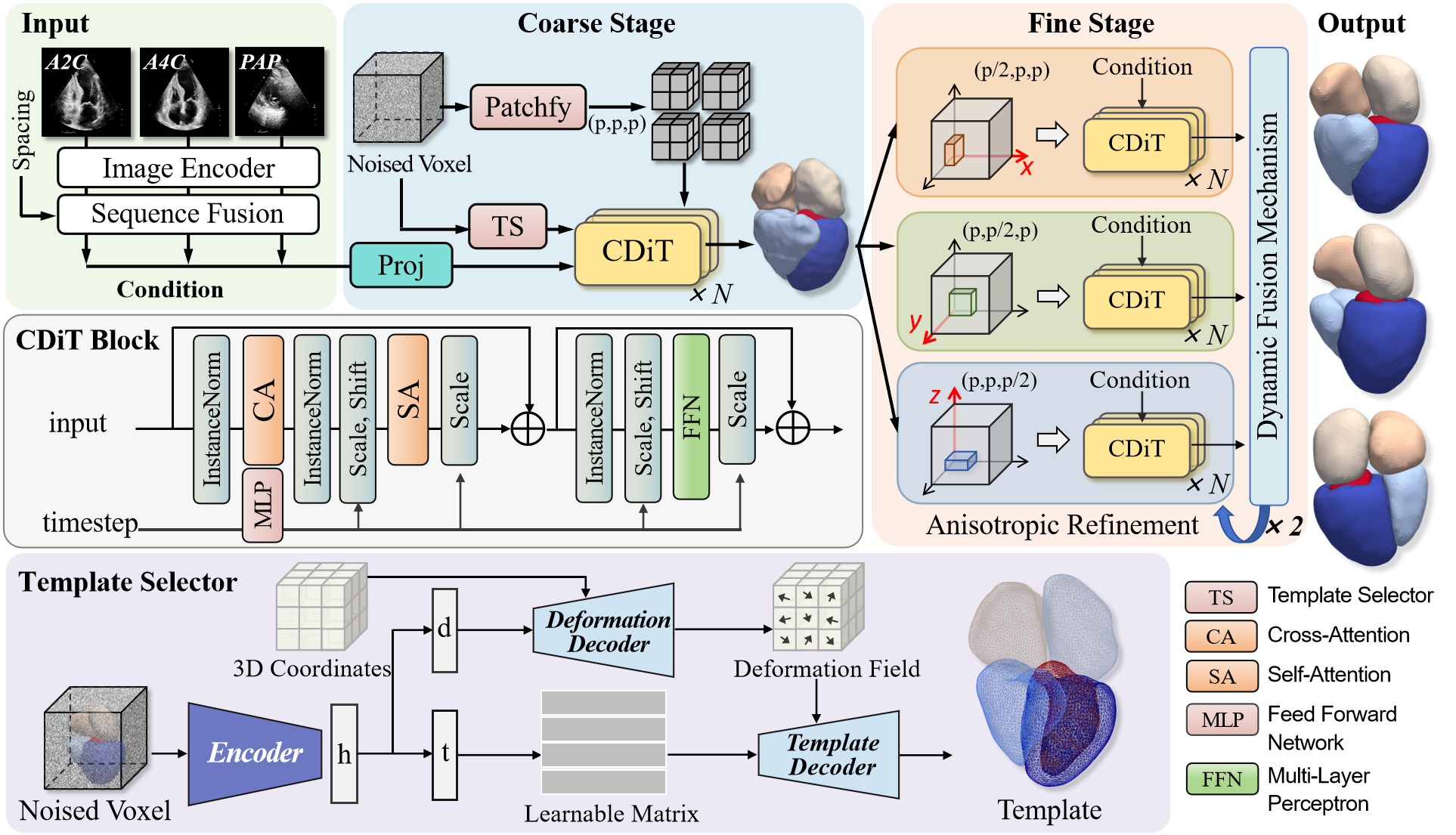}
\caption{The pipeline of our proposed UltraTwin. In the coarse stage, a template selector (TS) outputs a template to guide denoising. The coarse output is patchified and refined anisotropically, with double dynamic fusion of branch results.} 
\label{fig2}
\end{figure}

As shown in Fig.~\ref{fig2}, we propose UltraTwin to reconstruct personalized, high-fidelity 3D cardiac anatomical twins from sparse multi-view 2D US images under topological constraints. Our method combines pseudo-paired data pre-training, strictly paired data fine-tuning, a coarse-to-fine denoising scheme, and an implicit autoencoder for topology-aware reconstruction. Our baseline uses only coarse denoising scheme with condition injection.

\subsection{High-quality Dataset Construction}
To address the challenge of scarce paired data, we construct a high-quality dataset comprising both paired and unpaired multi-view 2D US images and 3D cardiac models. The unpaired data is processed using a novel Pseudo-paired Real Data Generation Strategy to enable effective pre-training.

\textbf{Paired Data Collection.}
The paired data includes multi-view 2D US images and ECG-gated CT scans. 
We use TotalSegmentator~\cite{wasserthal2023totalsegmentator} to segment cardiac structures from CT scans, extracting 3D models at end-diastole (ED) and end-systole (ES) phases.
For multi-view 2D US videos, ED and ES frames are manually annotated and temporally aligned, resulting in a strictly paired dataset.

\textbf{Pseudo-paired Real Data Generation Strategy.}
To address the scarcity of paired data, we leverage temporal-unpaired multi-view 2D US videos and chest CT scans to create additional training data for model pre-training. 
For each patient, the standard Apical 4-Chamber (A4C) view of 3D cardiac model is computed following~\cite{stojanovski2022efficient}. Seven key parameters are calculated on this view.
For US videos, pretrained structure and keypoint detectors extract cardiac motion curves and frame-level parameters. 
The A4C US frame with minimal parameter error compared to the 3D cardiac model is identified, and multi-view images in videos are temporally aligned based on their frame position in the motion curve. 

\subsection{Coarse-to-fine Conditional Diffusion Transformer}
To generate high-fidelity 3D cardiac models, we employ a coarse-to-fine conditional Diffusion Transformer (DiT), leveraging multiview feature fusion, cardiac template priors, and 
anisotropic refinement.

\textbf{Preliminaries.}
Following the original DiT-3D~\cite{mo2023dit}, the input voxel tensor \( V_i \in \mathbb{R}^{N \times V \times V \times V} \) is divided into patches of size \( (p, p, p) \), producing patch tokens \( t \in \mathbb{R}^{L \times N} \), where \( L = (V / p)^3 \) is total number of patches. 
In our study, \( N = 6 \) denotes the number of classes (5 cardiac structures and one background). 
We refer readers to DiT-3D~\cite{mo2023dit} for more details.
To effectively utilize multi-view 2D US images, we introduce a Multi-view Feature Fusion module and a Conditional DiT (CDiT)-Block. 
A pre-trained ResNet50~\cite{he2016deep} extracts view-level features, which are flattened into 2D vectors \( R \in \mathbb{R}^{d_I \times (h_f \cdot w_f)} \). 
We further inject view-level spacing information into the features using an MLP. 
As shown in Fig.~\ref{fig2}, these features are concatenated and fused via 1D convolutional blocks, then projected into a condition vector \( C \in \mathbb{R}^{L \times D} \). 
And the CDiT-Block extends the DiT-Block with a cross-attention mechanism for condition injection. 
The input patch embeddings \( e_v \in \mathbb{R}^{L \times D} \) are used as the query (\( q \)), while the condition vector \( C \) serves as the key (\( k \)) and value (\( v \)). The cross-attention operation (CA) is defined as
$\text{softmax}\left(\frac{qk^T}{\sqrt{d_k}}\right)v$.

\textbf{In the coarse stage}, we introduce cardiac template priors to guide the denoising process. 
The input voxels are fed into a template selector module, which comes from the pre-trained implicit autoencoder (details refer to Sec.~\ref{sec:2.3}) to select the most similar explicit cardiac template \( T \in \mathbb{R}^{N \times V \times V \times V} \). 
The template prior is patchified and embedded with positional encoding to obtain \( e_T \). 
With \( e_v \) as the query, \( e_T \) as key and value, and \( C \) as redisual, respectively, dual-condition cross-attention is applied for integration. 
This process repeats across multiple CDiT-Blocks, generating a coarse voxel tensor \( V_c \in \mathbb{R}^{N \times V \times V \times V} \). 

\textbf{In the refinement stage}, we propose an anisotropic patch partitioning and dynamic fusion strategy.
Our strategy includes the following key features. 
1) \textit{Computational Efficiency}: 
we anisotropicly partition in different branches (i.e.,\( (p/2, p, p) \), \( (p, p/2, p) \), \( (p, p, p/2) \)), to mitigate the high computational load of traditional uniform patch partitioning \((p/2, p/2, p/2)\), caused by extensive attention mechanisms in CDiT-Blocks.
2) \textit{Multi-directional Optimization}: 
each branch optimizes along a specific axis (X/Y/Z), capturing fine-grained details in different directions. 
The double fusion operation facilitates cross-branch information merging, enhancing global consistency. 
3) \textit{Dynamic Fusion Mechanism}: 
based on local variance (\( \sigma^2 \)) for voxel tensor ($V$), the normalized adaptive weight ($\omega_{x,y,z}$) fusion prioritizes low uncertainty branch outputs in the final result: 
    \begin{equation}
        \sigma^2 = \text{E}(V^2) - [\text{E}(V)]^2, \label{eq:fusion} 
    \end{equation}
     \begin{equation}
        w_x = \frac{1}{\sigma_x^2 + \epsilon}, \quad w_y = \frac{1}{\sigma_y^2 + \epsilon}, \quad w_z = \frac{1}{\sigma_z^2 + \epsilon},
    \end{equation}
    where \( \text{E}(V) \) is the local mean obtained via 3D average pooling. 

\subsection{Topology Constrained Implicit AutoEncoder}
\begin{figure}[!t]
\centering\includegraphics[width=1\textwidth]{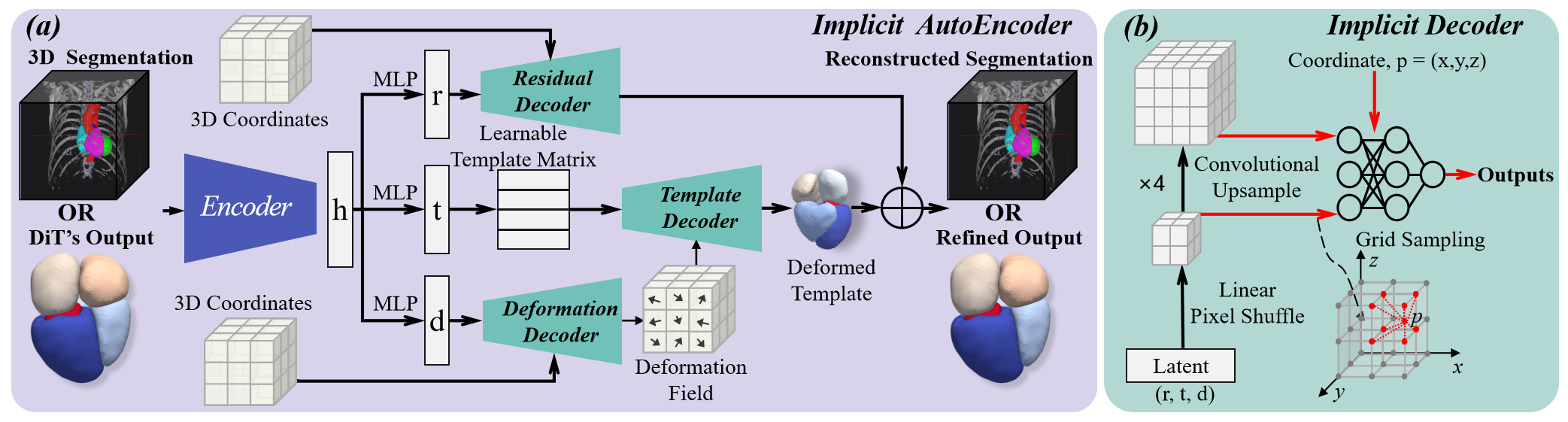}
\caption{Topology Constrained Implicit AutoEncoder. (a) The Pipeline of  Implicit AutoEncoder. (b) The details of Implicit Decoder.} 
\label{fig3}
\end{figure}

Reconstructing 3D cardiac models directly from US images is challenging due to inherent noise and artifacts. 
Direct denoising may yield topologically implausible results. 
To address this, we introduce a trained Implicit AutoEncoder during inference, imposing topological constraints on the conditional DiT's outputs to ensure accurate and plausible reconstructions.

\label{sec:2.3}
The autoencoder, trained on a large dataset of 3D segmentation from cardiac CT, compresses inputs into latents via multiple convolutional blocks. As shown in Fig.~\ref{fig3}, inspired by Yang et al.~\cite{yang2024generating}, it contains three decode branches: \textbf{Deformation (d)}, \textbf{Template (t)}, and \textbf{Residual (r)}. Each branch includes an MLP and a convolutional implicit decoder, with the T branch incorporating a learnable template matrix. The implicit decoder uses a linear pixel shuffle layer followed by four TrilinearUpsampleBlocks. Multi-scale features are extracted via grid sampling, concatenated, and fed into an MLP to predict point occupancy.

During training, the autoencoder is exposed to data corruption (noise addition, random masking, voxel value swapping) to enhance robustness. 
At inference, it refines the conditional DiT's outputs by enforcing anatomical constraints. The refined result is dynamically fused with the original output using spatially adaptive weighting (see Equation \eqref{eq:fusion}), yielding a topology-aware 3D cardiac model that balances detail preservation and anatomical accuracy.

\section{Experiments}
\textbf{Data Acquisition.} 
We prospectively collected multi-center (9 hospitals) US multi-view videos (12 standard views) paired with CT scans, with <10 days between examinations. 
The CT scans were categorized as ECG-gated (precisely paired) or non-contrast (unable to establish accurate 2D-3D pairing).
3D cardiac models were derived from CT segmentations using TotalSegmentator~\cite{wasserthal2023totalsegmentator}, with manually correction.
These models were resampled to 3mm spacing.
Then, they were rotated, decentralized, and transformed into 64$^{3}$ voxel data for training/testing. 
US images were resized to 224×224, with spacing updated. 
We collected 891 paired cases/patients: 96 ECG-gated and 795 non-strictly paired. 
For multi-view reconstruction, 130 strictly paired samples (since some cases include both ED/ES phases) were split at the patient level for training (96), validation (10), and testing (24).
Other cases were used for pre-training. 
For implicit autoencoder training, we used 1,379 cardiac models (excluding poor quality/incomplete segmentation), including 454 from TotalSegmentator~\cite{wasserthal2023totalsegmentator} and 891 from our dataset (excluding validation/test samples).
Only three standard views (A2C, A4C, PSAX\_PAP) were used for reconstruction.

\textbf{Implementation details.} 
All models were implemented in PyTorch and trained on one NVIDIA GTX 4090 GPU (24GB memory). 
The framework was trained for 400 epochs without pre-training (learning rate, $lr$=7e-5).
Alternatively, it was pre-trained for 200 epochs with $lr$=1e-4.
For DiT, we employed multi-class DSC loss with process supervision applied to outputs at multiple stages, using consistent weight coefficients. 
The loss function for implicit autoencoder followed Yang et al.~\cite{yang2024generating}. 
We kept all settings of other methods consistent with their original papers. 
Adam optimizer was used with batch size=4.
Models with best performance on the validation set were saved for testing.
Dataset and code are available at \url{https://github.com/JacksonYu-321/UltraTwin}.

\begin{table}[!h]
\centering
\caption{Quantitative comparison. Metrics include (DSC, HD, $E_{vol}$). Results are reported for both (w/o Pretrained) and (w Pretrained). Statistical significance is calculated on DSC against UltraTwin (w Pretrained). Acronyms: Myocardium (Myo), Left and Right Ventricle (LV, RV), Left and Right Atrium (LA, RA).}
\resizebox{\textwidth}{!}{%
\begin{tabular}{cc|cccccc|c|c|c}
\hline
\multicolumn{2}{c|}{\multirow{2}{*}{\textbf{Method}}} &
  \multicolumn{6}{c|}{\textbf{DSC} (\%)} &
  \textbf{HD} &
  \textbf{E$_{vol}$} &
  \textbf{p-value} \\ \cline{3-11} 
\multicolumn{2}{c|}{} &
  Myo &
  LV &
  LA &
  RV &
  \multicolumn{1}{c|}{RA} &
  Average &
  Average &
  Average &
  \multicolumn{1}{l}{(vs. UltraTwin)} \\ \hline
\multicolumn{1}{c|}{\multirow{8}{*}{\begin{tabular}[c]{@{}c@{}}w/o \\ pretrained\end{tabular}}} &
  E-Pix2Vox++\cite{stojanovski2022efficient} &
  65.20 &
  74.72 &
  74.33 &
  75.26 &
  \multicolumn{1}{c|}{73.43} &
  72.59 &
  6.59 &
  27.88 &
  \textless 0.001 \\
\multicolumn{1}{c|}{} &
  GARNet\cite{zhu2023garnet} &
  63.58 &
  80.71 &
  77.73 &
  71.56 &
  \multicolumn{1}{c|}{72.09} &
  73.13 &
  6.28 &
  24.36 &
  \textless 0.001 \\
\multicolumn{1}{c|}{} &
  UMIFormer\cite{zhu2023umiformer} &
  63.19 &
  80.96 &
  75.57 &
  76.50 &
  \multicolumn{1}{c|}{70.05} &
  73.25 &
  5.42 &
  30.38 &
  \textless 0.001 \\
\multicolumn{1}{c|}{} &
  LRGT\cite{yang2023long} &
  57.77 &
  78.84 &
  68.19 &
  72.42 &
  \multicolumn{1}{c|}{67.51} &
  68.95 &
  8.10 &
  34.07 &
  \textless 0.001 \\
\multicolumn{1}{c|}{} &
  Baseline &
  65.70 &
  81.67 &
  77.56 &
  76.92 &
  \multicolumn{1}{c|}{71.63} &
  74.69 &
  6.09 &
  23.29 &
  \textless 0.001 \\
\multicolumn{1}{c|}{} &
  +Implicit\_AE &
  \textbf{66.52} &
  82.36 &
  78.48 &
  77.90 &
  \multicolumn{1}{c|}{72.70} &
  75.59 &
  5.21 &
  24.95 &
  0.0229 \\
\multicolumn{1}{c|}{} &
  +Coarse-to-fine &
  65.90 &
  82.11 &
  78.67 &
  77.68 &
  \multicolumn{1}{c|}{73.80} &
  75.63 &
  5.31 &
  \textbf{22.14} &
  0.0269 \\
\multicolumn{1}{c|}{} &
  UltraTwin &
  66.48 &
  \textbf{82.52} &
  \textbf{79.10} &
  \textbf{78.16} &
  \multicolumn{1}{c|}{\textbf{74.18}} &
  \textbf{76.09} &
  \textbf{5.00} &
  23.27 &
  0.1355 \\ \hline
\multicolumn{1}{c|}{\multirow{8}{*}{\begin{tabular}[c]{@{}c@{}}w \\ pretrained\end{tabular}}} &
  E-Pix2Vox++\cite{stojanovski2022efficient} &
  63.36 &
  80.75 &
  76.70 &
  78.51 &
  \multicolumn{1}{c|}{74.21} &
  74.71 &
  5.54 &
  25.43 &
  \textless 0.001 \\
\multicolumn{1}{c|}{} &
  GARNet\cite{zhu2023garnet} &
  65.04 &
  82.67 &
  79.01 &
  78.22 &
  \multicolumn{1}{c|}{72.40} &
  75.47 &
  5.60 &
  22.60 &
  0.0138 \\
\multicolumn{1}{c|}{} &
  UMIFormer\cite{zhu2023umiformer} &
  63.20 &
  80.91 &
  75.52 &
  76.49 &
  \multicolumn{1}{c|}{70.09} &
  73.24 &
  5.36 &
  30.43 &
  \textless 0.001 \\
\multicolumn{1}{c|}{} &
  LRGT\cite{yang2023long} &
  59.12 &
  78.48 &
  74.77 &
  73.84 &
  \multicolumn{1}{c|}{67.31} &
  70.70 &
  5.89 &
  35.76 &
  \textless 0.001 \\
\multicolumn{1}{c|}{} &
  Baseline &
  66.66 &
  83.10 &
  78.10 &
  78.27 &
  \multicolumn{1}{c|}{74.11} &
  76.05 &
  5.27 &
  22.99 &
  0.0366 \\
\multicolumn{1}{c|}{} &
  +Implicit\_AE &
  67.27 &
  83.24 &
  79.22 &
  79.19 &
  \multicolumn{1}{c|}{74.61} &
  76.70 &
  5.09 &
  24.19 &
  0.2897 \\
\multicolumn{1}{c|}{} &
  +Coarse-to-fine &
  67.01 &
  83.24 &
  79.40 &
  78.38 &
  \multicolumn{1}{c|}{75.52} &
  76.71 &
  5.16 &
  \textbf{21.33} &
  \textless 0.001 \\
\multicolumn{1}{c|}{} &
  UltraTwin &
  \textbf{67.82} &
  \textbf{83.66} &
  \textbf{79.88} &
  \textbf{79.21} &
  \multicolumn{1}{c|}{\textbf{75.80}} &
  \textbf{77.27} &
  \textbf{5.07} &
  22.56 &
  - \\ \hline
\end{tabular}%
}
\label{tab:results}
\end{table}

\textbf{Evaluation Metrics}.
The quantitative evaluation was performed using well-established cardiac imaging metrics, i.e., the Dice Similarity Coefficient (DSC) to assess spatial overlap accuracy, the Hausdorff Distance (HD) to quantify contour alignment, and volumetric error ($E_{vol}$, ml) to evaluate anatomical consistency.

\textbf{Quantitative Comparisons.}
Our proposed UltraTwin consistently outperforms all competing methods across multiple structures and metrics ($w$ or $w/o$ pretraining), achieving statistically significant improvements ($p<0.05$, Wilcoxon test).
Without pretraining setting, UltraTwin achieves the highest average DSC (76.09), the lowest HD (5.00), and competitive $E_{vol}$ (23.27 ml). 
With pretraining, UltraTwin further improves, reaching an average DSC of 77.27, HD of 5.07, and $E_{vol}$ of 22.56 ml. 
These results demonstrate the effectiveness of our approach.
Besides, pretraining with our proposed pseudo-paired data significantly enhances the performance of all methods, highlighting its effectiveness in bridging the domain gap and improving reconstruction accuracy.

\textbf{Ablation Studies.} 
We test the contributions of key components in UltraTwin. 
The Implicit Autoencoder (Implicit$\_\text{AE}$) enforces topological and anatomical constraints, refining the initial reconstruction to produce clinically plausible cardiac structures and improving the average DSC from 76.05 to 76.70. 
Besides, the coarse-to-fine strategy achieves satisfactory DSC and HD, while obtaining the lowest $E_{vol}$ (21.33 ml). 
Combining all components achieves state-of-the-art performance, except for $E_{vol}$, as the topological constraints prioritize generating topologically consistent shapes over exact volume matching.

\begin{figure}[!t]
\centering
\includegraphics[width=1\textwidth]{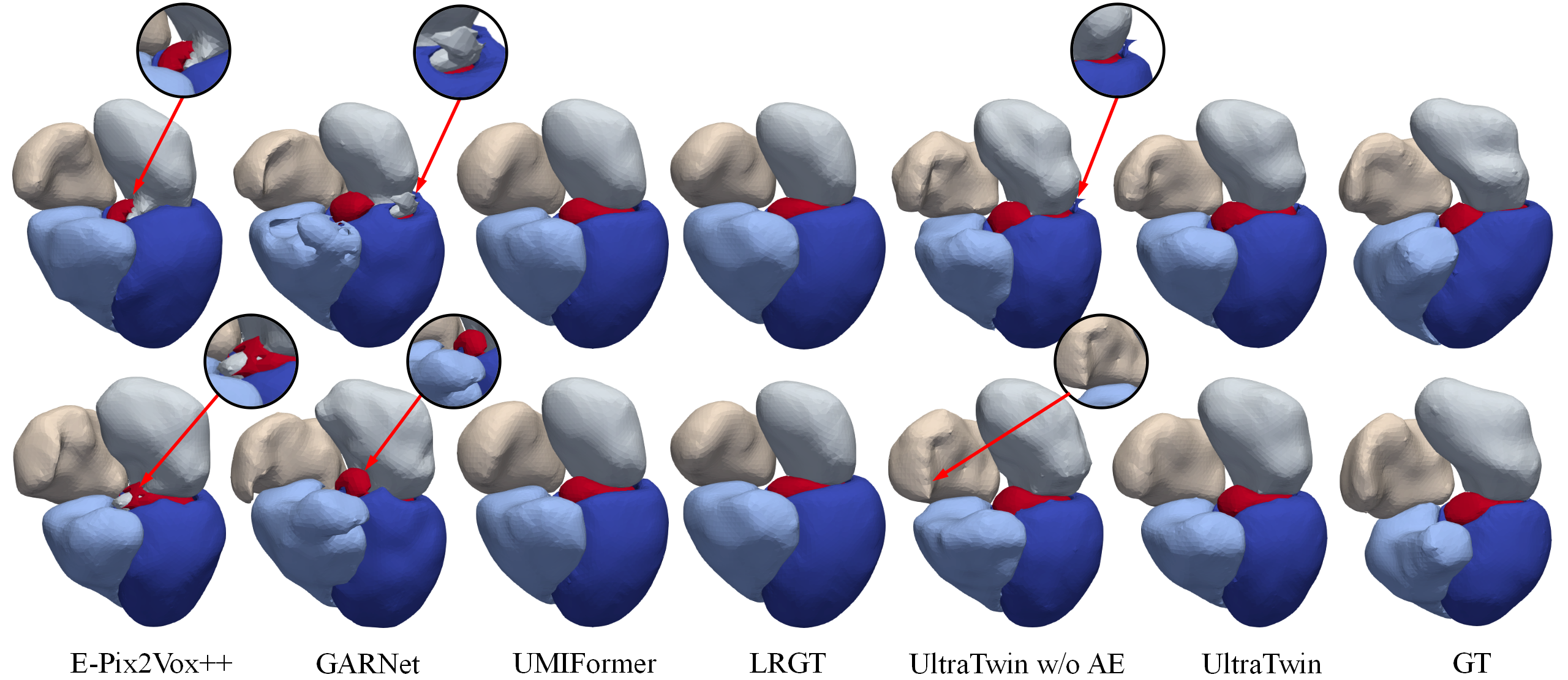}
\caption{Visual comparison of different methods (fine-tuned). Rows 1 and 2 represent the ED and ES phase of the same patient, respectively.}
\label{fig4}
\end{figure}

\begin{figure}[!t]
\centering
\includegraphics[width=1\textwidth]{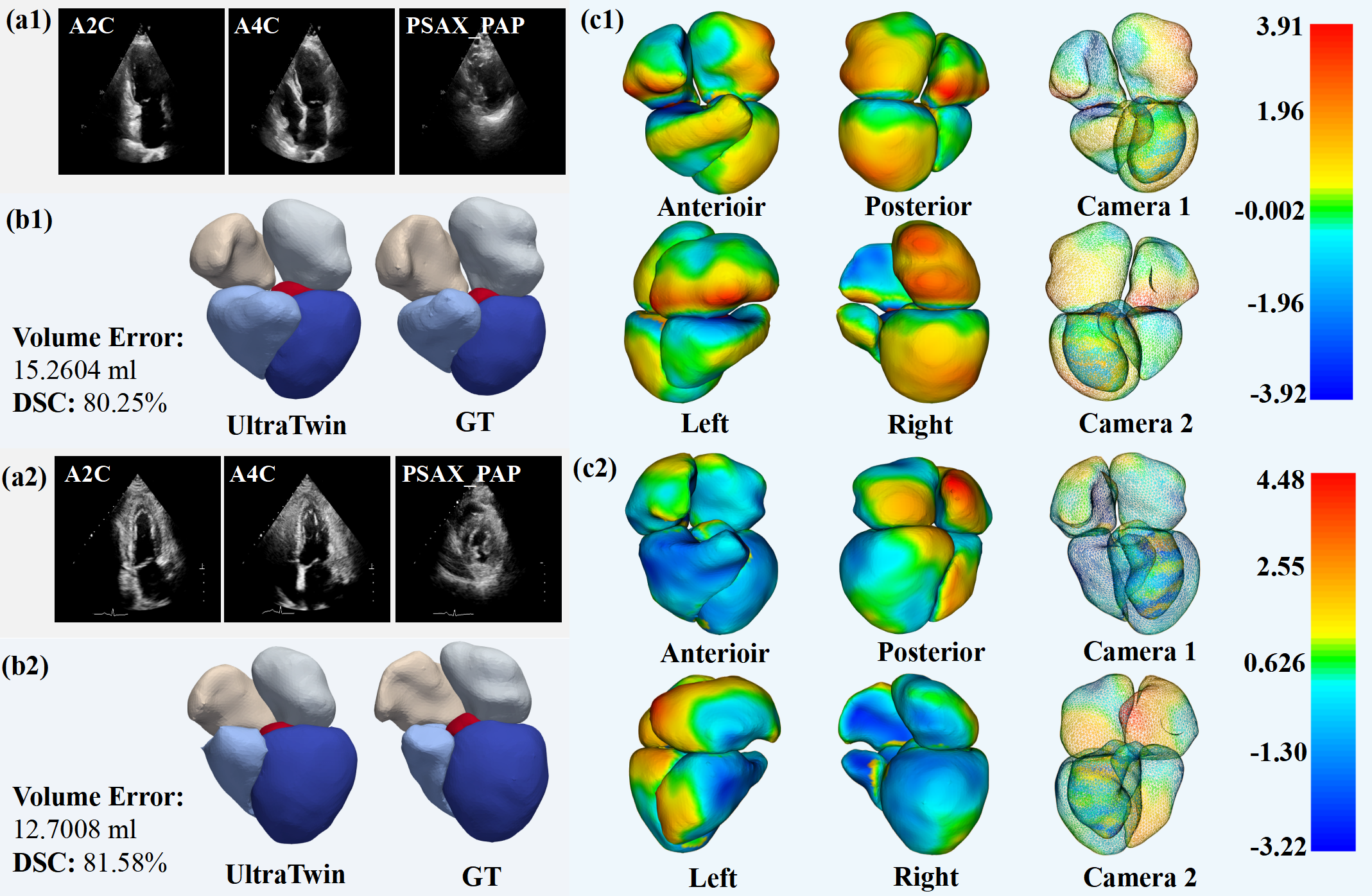}
\caption{Visual comparison between UltraTwin and GT. 
(a1, b1, c1) and (a2, b2, c2) represent the input US images, reconstructed cardiac digital twins, GTs, and surface distance maps from different camera views for two patients, respectively.}
\label{fig6}
\end{figure}

\textbf{Qualitative analysis.}
As shown in Fig.~\ref{fig4}, transformer-based UMIFormer and LRGT exhibit model collapse, reconstructing nearly identical results despite competitive quantitative metrics (see Table~\ref{tab:results}). 
We hypothesize that transformer-based methods may struggle with limited data and reconstruction ambiguity. 
In contrast, UltraTwin produces personalized and topologically plausible anatomical twins, mainly driven by our proposed topological constraints. 
Specifically, these constraints refine the initial reconstruction by deforming implausible structures.
For example, irregular protrusions or discontinuities in the initial reconstruction are corrected, producing smoother and more realistic anatomical twins.
Fig.~\ref{fig6} visualizes the reconstruction of UltraTwin for two cases, comparing with GTs.
The comparison demonstrates that UltraTwin generates highly accurate and personalized cardiac anatomical twins.

\section{Conclusion}
We propose UltraTwin, a conditional generative model, to construct cardiac anatomical twin from sparse multi-view 2D US images.
We first construct a high-quality real-world dataset with rigorously paired and pseudo-paired data.
Then, a coarse-to-fine scheme was introduced to achieve anisotropic and progressive refinement. 
Last, we incorporate an implicit autoencoder for topology-aware constraints during inference. 
UltraTwin beats strong competitors in both quantitative and qualitative evaluations. 
In the future, we will extend UltraTwin for personalized cardiac function evaluation and optimization of treatment planning.

\begin{credits}
\subsubsection{\ackname} This work was supported by the grant from National Natural Science Foundation of China (82201851, 12326619, 62171290); Guangxi Province Science Program (2024AB17023); Yunnan Major Science and Technology Special Project Program (202402AA310052); Yunnan Key Research and Development Program (202503AP 140014); Suzhou Gusu Health Talent Program (GSWS 2022071); Suzhou Applied Basic Research (Medical and Health) Technology Innovation Projects (SYWD2024009); Science and Technology Planning Project of Guangdong Province (2023A0505020002); Frontier Technology Development Program of Jiangsu Province (BF2024078).

\subsubsection{\discintname}
The authors have no competing interests to declare that are relevant to the content of this article.
\end{credits}

\bibliographystyle{splncs04}
\bibliography{Paper-0875}

\begin{thebibliography}{10}
\providecommand{\url}[1]{\texttt{#1}}
\providecommand{\urlprefix}{URL }
\providecommand{\doi}[1]{https://doi.org/#1}

\bibitem{bernard2018deep}
Bernard, O., Lalande, A., Zotti, C., Cervenansky, F., Yang, X., Heng, P.A., Cetin, I., Lekadir, K., Camara, O., Ballester, M.A.G., et~al.: Deep learning techniques for automatic mri cardiac multi-structures segmentation and diagnosis: is the problem solved? IEEE transactions on medical imaging  \textbf{37}(11),  2514--2525 (2018)

\bibitem{he2016deep}
He, K., Zhang, X., Ren, S., Sun, J.: Deep residual learning for image recognition. In: Proceedings of the IEEE conference on computer vision and pattern recognition. pp. 770--778 (2016)

\bibitem{dm}
Ho, J., Jain, A., Abbeel, P.: Denoising diffusion probabilistic models. Advances in neural information processing systems  \textbf{33},  6840--6851 (2020)

\bibitem{laumer2023weakly}
Laumer, F., Amrani, M., Manduchi, L., Beuret, A., Rubi, L., Dubatovka, A., Matter, C.M., Buhmann, J.M.: Weakly supervised inference of personalized heart meshes based on echocardiography videos. Medical image analysis  \textbf{83},  102653 (2023)

\bibitem{laumer20252d}
Laumer, F., Rubi, L., Matter, M.A., Buoso, S., Fringeli, G., Mach, F., Ruschitzka, F., Buhmann, J.M., Matter, C.M.: 2d echocardiography video to 3d heart shape reconstruction for clinical application. Medical Image Analysis  \textbf{101},  103434 (2025)

\bibitem{li2024towards}
Li, L., Camps, J., Wang, Z., Beetz, M., Banerjee, A., Rodriguez, B., Grau, V.: Towards enabling cardiac digital twins of myocardial infarction using deep computational models for inverse inference. IEEE transactions on medical imaging  (2024)

\bibitem{mo2023dit}
Mo, S., Xie, E., Chu, R., Hong, L., Niessner, M., Li, Z.: Dit-3d: Exploring plain diffusion transformers for 3d shape generation. Advances in neural information processing systems  \textbf{36},  67960--67971 (2023)

\bibitem{nelson2000sources}
Nelson, T., Pretorius, D., Hull, A., Riccabona, M., Sklansky, M., James, G.: Sources and impact of artifacts on clinical three-dimensional ultrasound imaging. Ultrasound in Obstetrics and Gynecology: The Official Journal of the International Society of Ultrasound in Obstetrics and Gynecology  \textbf{16}(4),  374--383 (2000)

\bibitem{peebles2023scalable}
Peebles, W., Xie, S.: Scalable diffusion models with transformers. In: Proceedings of the IEEE/CVF international conference on computer vision. pp. 4195--4205 (2023)

\bibitem{stojanovski2022efficient}
Stojanovski, D., Hermida, U., Muffoletto, M., Lamata, P., Beqiri, A., Gomez, A.: Efficient pix2vox++ for 3d cardiac reconstruction from 2d echo views. In: International Workshop on Advances in Simplifying Medical Ultrasound. pp. 86--95. Springer (2022)

\bibitem{thangaraj2024cardiovascular}
Thangaraj, P.M., Benson, S.H., Oikonomou, E.K., Asselbergs, F.W., Khera, R.: Cardiovascular care with digital twin technology in the era of generative artificial intelligence. European Heart Journal  \textbf{45}(45),  4808--4821 (2024)

\bibitem{wasserthal2023totalsegmentator}
Wasserthal, J., Breit, H.C., Meyer, M.T., Pradella, M., Hinck, D., Sauter, A.W., Heye, T., Boll, D.T., Cyriac, J., Yang, S., et~al.: Totalsegmentator: robust segmentation of 104 anatomic structures in ct images. Radiology: Artificial Intelligence  \textbf{5}(5),  e230024 (2023)

\bibitem{yang2024generating}
Yang, J., Sedykh, E., Adhinarta, J.K., Le, H., Fua, P.: Generating anatomically accurate heart structures via neural implicit fields. In: International Conference on Medical Image Computing and Computer-Assisted Intervention. pp. 264--274. Springer (2024)

\bibitem{yang2023long}
Yang, L., Zhu, Z., Lin, X., Nong, J., Liang, Y.: Long-range grouping transformer for multi-view 3d reconstruction. In: Proceedings of the IEEE/CVF International Conference on Computer Vision. pp. 18257--18267 (2023)

\bibitem{zhu2023umiformer}
Zhu, Z., Yang, L., Li, N., Jiang, C., Liang, Y.: Umiformer: Mining the correlations between similar tokens for multi-view 3d reconstruction. In: Proceedings of the IEEE/CVF International Conference on Computer Vision. pp. 18226--18235 (2023)

\bibitem{zhu2023garnet}
Zhu, Z., Yang, L., Lin, X., Yang, L., Liang, Y.: Garnet: Global-aware multi-view 3d reconstruction network and the cost-performance tradeoff. Pattern Recognition  \textbf{142},  109674 (2023)

\end{thebibliography}

\end{document}